\documentclass[journal]{vgtc}                

\ifpdf
  \pdfoutput=1\relax                   
  \usepackage{graphicx}                
  \DeclareGraphicsExtensions{.pdf,.png,.jpg,.jpeg} 
\else
  \usepackage{graphicx}                
  \DeclareGraphicsExtensions{.eps}     
\fi%

\graphicspath{{}{pictures/}{images/}{./}} 

\usepackage{microtype}                 
\PassOptionsToPackage{warn}{textcomp}  
\usepackage{textcomp}                  
\usepackage{mathptmx}                  
\usepackage{times}                     
\usepackage{cite}                      
\usepackage{tabu}                      
\usepackage{booktabs}                  

\usepackage{array} 
\newcolumntype{L}[1]{>{\raggedright\let\newline\\\arraybackslash\hspace{0pt}}p{#1}}
\newcolumntype{C}[1]{>{\centering\let\newline\\\arraybackslash\hspace{0pt}}p{#1}}
\newcolumntype{R}[1]{>{\raggedleft\let\newline\\\arraybackslash\hspace{0pt}}p{#1}}

\onlineid{1090}
\vgtccategory{theory/model}
\vgtcpapertype{theory/model}

\title{Data by Proxy --- Material Traces as Autographic Visualizations}

\author{Dietmar Offenhuber \emph{Northeastern University}} \authorfooter{

\item Dietmar Offenhuber is with Northeastern University. E-mail: d.offenhuber@northeastern.edu. }

\shortauthortitle{Offenhuber \MakeLowercase{\emph{et al.}}: Data by Proxy Autographic Visualization}

\abstract{Information visualization limits itself, per definition, to the domain of symbolic information. This paper discusses arguments why the field should also consider forms of data that are not symbolically encoded, including physical traces and material indicators. Continuing a provocation presented by Pat Hanrahan in his 2004 IEEE Vis capstone address, this paper compares physical traces to visualizations and describes the techniques and visual practices for producing, revealing, and interpreting them. By contrasting information visualization with a speculative counter model of autographic visualization, this paper examines the design principles for material data. Autographic visualization addresses limitations of information visualization, such as the inability to directly reflect the material circumstances of data generation. The comparison between the two models allows probing the epistemic assumptions behind information visualization and uncovers linkages with the rich history of scientific visualization and trace reading. The paper begins by discussing the gap between data visualizations and their corresponding phenomena and proceeds by investigating how material visualizations can bridge this gap. It contextualizes autographic visualization with paradigms such as data physicalization and indexical visualization and grounds it in the broader theoretical literature of semiotics, science and technology studies (STS), and the history of scientific representation. The main section of the paper proposes a foundational design vocabulary for autographic visualization and offers examples of how citizen scientists already use autographic principles in their displays, which seem to violate the canonical principles of information visualization but succeed at fulfilling other rhetorical purposes in evidence construction. The paper concludes with a discussion of the limitations of autographic visualization, a roadmap for the empirical investigation of trace perception, and thoughts about how information visualization and autographic visualization techniques can contribute to each other.} 


\CCScatlist{ 

\CCScat{K.6.1}{Human-centered computing}
{Project and People Management}{Life Cycle}; \CCScat{K.7.m}{The Computing Profession}{Miscellaneous}{Ethics} }

\teaser{ 
\centering 
\includegraphics[width=\linewidth]{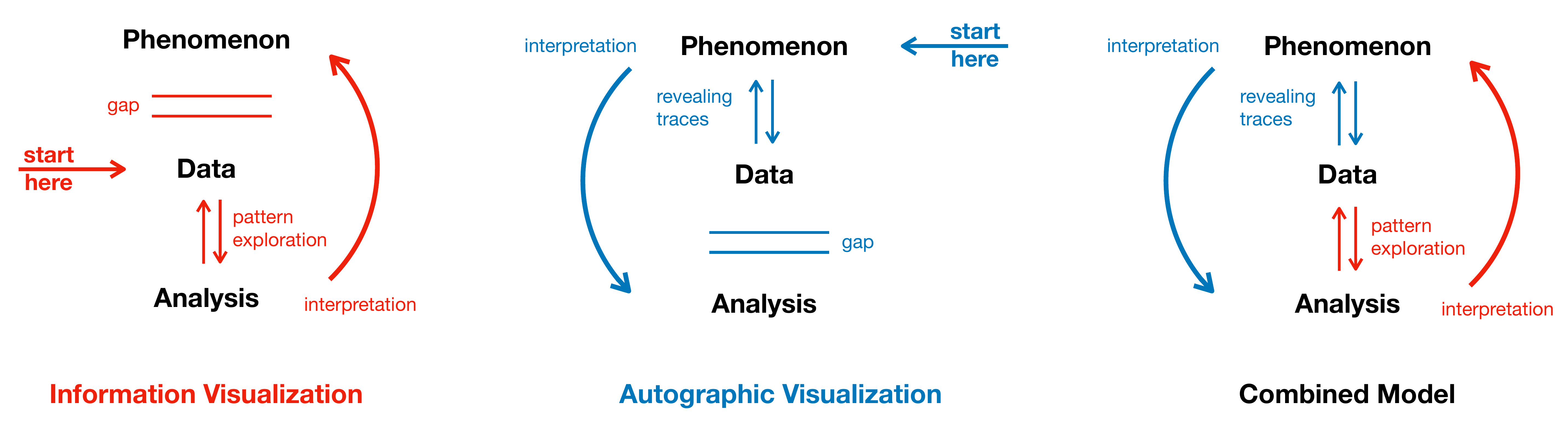} 
\caption{Traditional information visualization starts with the exploration of a data set to find inherent patterns. The represented phenomenon is interpreted based on what is found in the data, but the connection between data and phenomenon remains hidden. Autographic visualization starts with the phenomenon and explores the data generation process through material traces. A gap remains between the interpretation of traces and more complex forms of computational analysis. A combined model uses autographic principles of revealing traces to re-contextualize data with the phenomena they supposedly describe.}\label{fig:sketch}}

\vgtcinsertpkg{}

\begin{document}

\firstsection{Limitations of Symbolic Representation}

\maketitle

Information visualization is, by definition, bound to the domain of symbolic representation---information encoded in numbers and letters. 
Symbolic abstraction offers many advantages, including generalizable visual methods for pattern discovery and visual computation. 
However, visualization can only begin when data already exist. As a result, the material processes and circumstances of data collection remain largely hidden. Without additional context, a falsified data set may be indistinguishable from an authentic one. InfoVis proceeds to further abstract an already reductive set of observations and investigates external phenomena by looking inwards: seeking patterns and inconsistencies in the data sets representing these phenomena. Data and phenomenon, however, are separated by a gap that obscures their relationship and shared history (Fig.~\ref{fig:sketch}, left).

There are, however, situations where the circumstances of data generation are of central interest and subject to public controversy. In the case of climate change and environmental pollution, for example, it is not just the discovery of hidden patterns that matter, but the question of how even the most obvious patterns relate to the physical world. Since both climate and pollution are  statistical concepts based on long-term averages and threshold values, data visualization seems to be the obvious mode of representation~\cite{Edwards2013vast,Schneider2018Klimabilder}. Yet, the mantra of “above all else show the data”~\cite{Tufte1983Visual} is complicated by the fact that data gains meaning only through context and that data and context cannot always be distinguished. The visual languages of information visualization, however, tend to further decontextualize data for the sake of pattern discovery. Climate change skeptics often circulate charts based on data from ice cores and other paleoclimatic proxies to support arguments for “global cooling,” yet showing data accurately and providing the correct references. They mislead by foregrounding patterns that cannot be correctly interpreted without a deeper understanding of paleoclimatology and its conventions. The obvious response to this issue is that complex matters require more complex explanations and information displays. After all, dealing with complexity is considered one of the main strengths of information visualization. But this strength is of little help if it is the relationship between the data set and physical reality that is called into question.

To address the material circumstances of data collection, it is helpful to consider a broader definition of data. Paleoclimatologists use a vast array of data proxies, including tree rings, ice cores, and bioindicators such as plants, animals, and microbes~\cite{Mattern2017Big,Kuznetsov2011Nurturing}. These proxies are not just the raw material for creating data sets; we can also see them as physical forms of data. The narrow focus on symbolically encoded data in computer science is not universally shared across fields. Archaeologists describe the artifacts they extract from a field site as data, and also physicists frequently treat information as a material property. Philosopher Luciano Floridi defines a \emph{datum} as a “lack of uniformity,” echoing the cyberneticist Gregory Bateson who defined a bit of information as the “difference that makes a difference''~\cite{Floridi2011Philosophy,Bateson1972Steps}. The lack of uniformity includes any kinds of contrasts that manifest themselves long before any act of encoding.

\begin{figure}
	[ht] \centering 
	\includegraphics[width=3.45in]{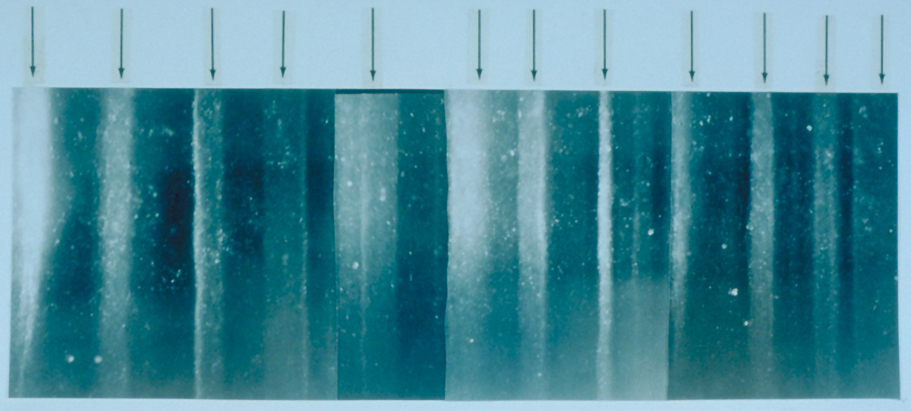} 
	\caption{GISP2 ice core section showing annual layer structure (cropped), illuminated from below. Source:~\url{ftp://ftp.ncdc.noaa.gov/pub/data/paleo/slidesets/icecore-polar/}}\label{ice}
\end{figure}

Ice cores and tree rings are not just material, but also visual forms of information that can be investigated through visual methods (Fig.~\ref{ice}). As conspicuous records of slow processes, their layers bear testament of past conditions. In this paper, we regard them as autographic visualizations: phenomena that reveal themselves as visible traces. Are all traces---and therefore almost anything in the world---autographic visualizations? A focus on practice avoids such a semantic dilution. Autographic visualization is a technique more than a thing: the application of specific methods to reveal environmental information as a trace, even if it involves only a particular observation skill. To observe tree rings, one has to cut into a tree and prepare the sample in specific ways. Beyond self-evident visual patterns, also non-visual qualities can be visualized, for example, by adding a tracer substance. Autographic visualization compares such material transformations to the process of visualizing data. 

\section{Autographic Visualization}
We present autographic visualization as a counter-model to data visualization, focusing on material rather than data encoded in numbers and symbols. It is a counter-model not in the sense that it excludes data visualization---there is considerable overlap between the two models---but in the sense that it clarifies the characteristics of each model through this juxtaposition. 

We refer to autographic visualization as a set of techniques for revealing material phenomena as visible traces and guiding their interpretation. Designing an autographic display means setting the conditions that allow a trace to emerge. We understand a trace as any transient or persistent configuration of matter presenting itself to the senses.

A central goal of autographic visualization is to make environmental information legible and the processes of data collection and their underlying causalities experiential and accountable. Since a data set is the outcome rather than the starting point (Fig.~\ref{fig:sketch}, middle), autographic visualization cannot rely on the representation of data. It is non-representational: rather than re-presenting something absent, the phenomenon presents itself. Autographic visualizations can be accidental, such as the desire paths across grass areas in the city or the uneven traces of wear on a staircase or a computer keyboard. But in general, they are the outcome of design operations that aim to reveal, isolate, amplify, conserve, and present material traces as records of past processes and events. For example, the display of the sundial~\cite{Peters2015Marvelous} is a product of a natural phenomenon untouched by human intervention. At the same time, it is a computational device designed to calculate not only the time of day but also month and season. Its display often incorporates a calendar---a classic form of data visualization---geometrically aligned with the path of the sun in the particular location.


Autographic visualization techniques draw from a long history of epistemic and material cultures that deal with the visual interpretation of traces, symptoms, or signatures as forms of material evidence. Its practices range from scientific experimentalism to ancient techniques of hunting, navigating, and healing. This paper is based on two premises. First, the diverse space of practices engaging with traces can be generalized into several distinct design operations. And second, these visual operations of autographic visualization are closely related to the modes of exploration facilitated by information visualization.

While the interpretation of medical symptoms, the design of experimental systems, or the design of shape-changing materials are usually considered in isolation, autographic visualization identifies common visual strategies across all of these practices. Table~\ref{infovsauto} summarizes the main differences between InfoVis and autographic visualization. To avoid confusion, we use \textit{symbolic data} to refer to digital data.

\begin{table}
	\caption{Comparison between InfoVis and Autographic Visualization}\label{infovsauto}\centering\scriptsize\medskip
	\begin{tabular}
		{L{1.8cm} L{2.7cm} L{3cm}} & InfoVis & Autographic Visualization\\
		\midrule Role of symbolic data & Begins with data & Ends with data\medskip\\
		Focus & Inwards: reveals patterns within data & Outwards: reveals the process of data generation\medskip\\
		Role of representation & Representational: visual marks stand for a phenomenon & Non-representational: the phenomenon presents itself\medskip\\
		Role of design & Mapping data to visual variables \& layouts & Elucidating qualities of a phenomenon\\
	\end{tabular}
\end{table}

Despite these differences, there is a close kinship between autographic and information visualization—both are rooted in the same visual culture and take advantage of similar perceptual mechanisms~\cite{Noe2004Action,Haroz2006Natural}. Foundational literature in visualization and HCI frequently invokes natural phenomena as metaphor or inspiration. Whether charts and graphs should be viewed as abstractions of natural phenomena based on shared organizational principles or as metaphorical references will not be elaborated here. However, it is worth noting that both InfoVis and autographic visualization were at one point considered to be the same approach. Etienne-Jules Marey's late 19\textsuperscript{th}~century \emph{Methode Graphique} encompasses both the charting of statistical information and the construction of self-registering devices for recording blood pressure, the flight of birds, or the turbulence of air. In pursuit of his declared goal to capture “the language of the phenomena themselves,” Marey's pioneering work included autographic devices such as the wind tunnel and, most prominently, his invention of chronophotography~\cite{Marey1878La}, inspiring other non-mimetic uses of photography~\cite{felton_photoviz:_2016}.

Analog information visualizations are often at the same time physical traces. Mechanically excited by seismic movements, a simple seismometer produces a classic line chart. This dual role creates a conceptual ambiguity that blurs the boundary between InfoVis and autographic visualization. If the line chart produced by the seismometer is a physical trace, what about a satellite image, what about the electrical charge generated by a digital sensor connected to a computer? The difference between an analog and a digital medium is not relevant for the underlying causality since both devices operate in a deterministic way. From this perspective, many symbolic datasets indeed share the character of a material trace; the material aspects of data collection inscribe themselves, sometimes unintentionally, into the data set~\cite{Dourish2013Media}. 

This can be illustrated through a public data set of GPS traces of drop-off and pick-up locations of NYC taxis. Plotting the data set in Cartesian space yields, unsurprisingly, a figure that resembles a map of the city. Some areas on this map, however, appear blurrier than others: an artifact of diminished GPS reception between tall buildings. In other words, the two-dimensional geographic datum contains hidden information about the three-dimensional shape of the city. But this latent information is only accessible if the materiality of GPS is understood and considered. A material reading that takes advantage of such artifacts, or “dust” in the data~\cite{Loukissas2016Taking}, differs from a classic approach of cleaning the data set by excluding obvious errors, e.g., points that fall into the ocean or within buildings. While information- and autographic visualization may differ in the length of the causal chains that link phenomenon and representation, the autographic perspective can to some extent also be applied to digital information, further explored in Section~\ref{performative}.

\begin{figure*}
	[t] \centering 
	\includegraphics[width=18cm]{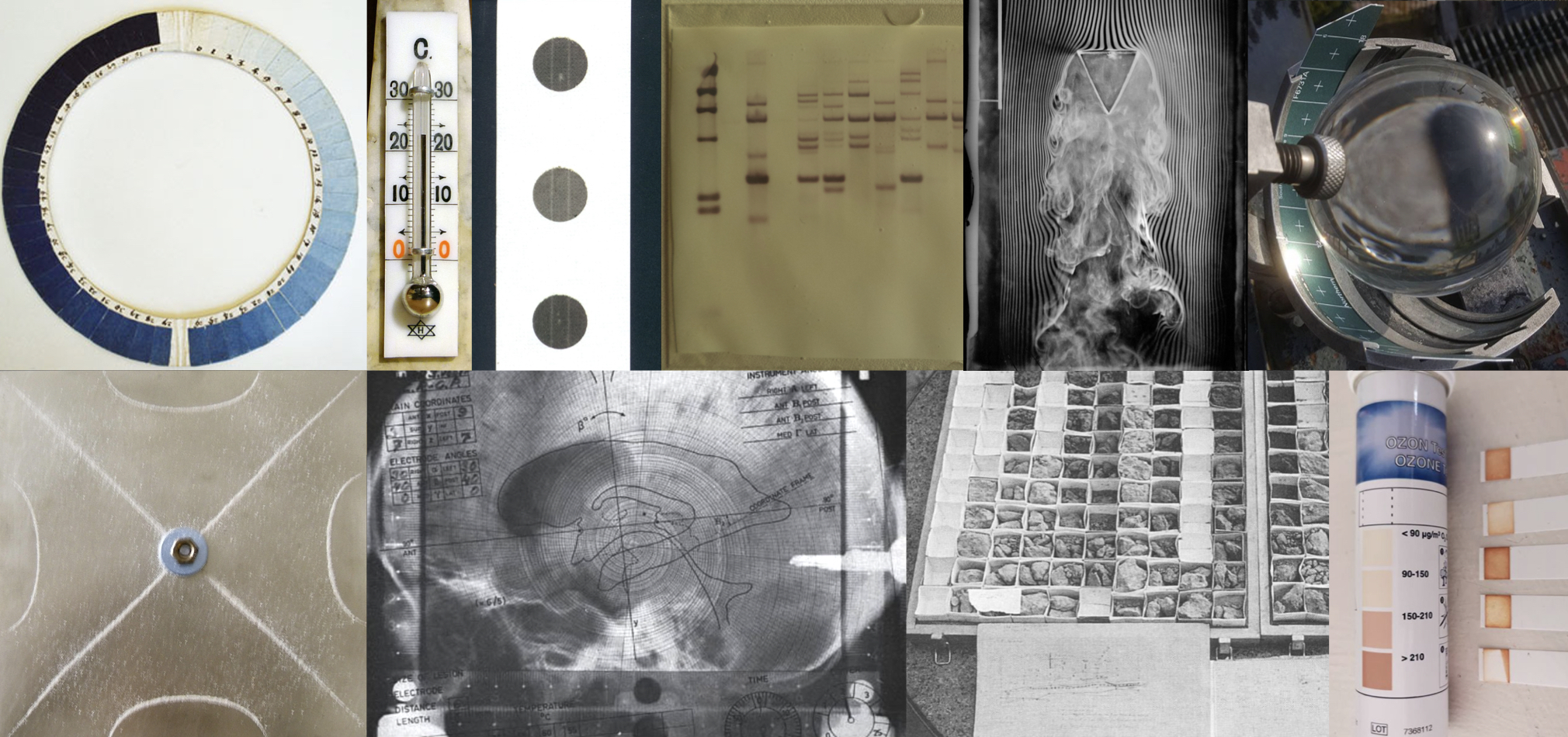} 
	\caption{Autographic visualizations and their design operations (Table~\ref{designop}), top-left to bottom-right: (a)~Cyanometer, a device for measuring the blueness of the sky, \emph{framing} and \emph{encoding}~\cite{Saussure1791Description}; (b)~Mercury-in-glass thermometer, \emph{constraining} and \emph{encoding}; (c)~filter for sampling airborne particulate matter, \emph{aggregating}; (d) southern blot for DNA electrophoresis, \emph{separating, registering}; (e)~EJ Marey’s smoke machine to visualize airflow, \emph{coupling}; (f)~Campbell-Stokes sunshine recorder, \emph{registering, encoding}; (g)~Chladni figure revealing sound waves, \emph{coupling, registering}; (h)~a planning diagram for neuro-surgery, \emph{annotating}~\cite{Wulfingen2017Traces:}; (i)~pedocomparator for sampling and comparing soil samples, \emph{aggregating, encoding}~\cite{Latour1999Pandora's}; (j)~reagent strips for ozone detection, \emph{registering, encoding}.}\label{collage}
\end{figure*}

\section{Theoretical Perspectives on Material Traces}
Classic semiology, in many ways foundational for information design and visualization~\cite{Bertin1983Semiology,Robinson1975Map,Cleveland1987Graphical}, offers a framework for analyzing physical traces. Philosopher Charles Sanders Peirce differentiates \emph{symbol}, \emph{icon}, and \emph{index} as three kinds of (non-exclusive) relationships a sign can have with a corresponding object or concept in the world. The symbol is linked to its object based on arbitrary convention; the icon is based on a relationship of resemblance; and in the case of the index, the relationship is an existential connection such as the causal link between a footprint and the person that left it~\cite{Peirce1998What}. While \emph{icons} (Peirce includes diagrams in this category) and \emph{symbols} play a prominent role in information design and visualization, indexical signs appear only implicitly; for example, as patterns and signals in data sets. 

Indexical phenomena have been explored in HCI, ubiquitous computing, and to a lesser extent, information visualization~\cite{Moere2009Analyzing,Offenhuber2012Kuleshov's,Schofield2013Indexicality,Offenhuber2015Indexical}. The application of indexicality to physical traces, however, is somewhat limited by the central role of linguistic concepts in semiotic theory. Peirce, for example, describes a pointing finger, a physical trace, and the word “there” as equivalent examples of indexical signs. By relying on the sign as the universal vehicle of meaning, semiotic perspectives reduce the trace to its role as a signifier. Scholars have critiqued the semiotic model of representation, in which meaning is conveyed by signs that stand for concepts in the world. To paraphrase historian Lorraine Daston, the proposition of a one-to-one correspondence between a sign and its object turned out to be as useless as the Borgesian 1-to-1 map that fully covers the territory~\cite{Coopmans2014Representation}.

When considering all the processes, actions, and material conditions involved in exploring traces, it is not always useful to make explicit what exactly constitutes a sign and how it is used to generate meaning. Scholars in science, technology, and society (STS) have formulated alternative perspectives that focus on the performative and embodied modes of cognition with regard to the roles of traces and trace-making in the history of science. Bruno Latour describes data, traces, and visualizations as \emph{immutable mobiles}: aspects of the world that have been stabilized, flattened, and made mobile to support arguments in scientific discourse~\cite{Latour1990Visualisation,Latour1999Pandora's}. In a similar vein, Hans-Jörg Rheinberger speaks about epistemic things: objects manipulated in the laboratory that should not just be regarded as samples collected from the world, but as materializations of research questions and scientific models that are embodied in the countless transformations applied to them~\cite{Rheinberger1997Toward}.

The notion of the trace as objective evidence and science as a process of trace-making has blossomed in the 19\textsuperscript{th}~century paradigm of \emph{mechanical objectivity}. Charting the history of objectivity through scientific atlases and visualization, historians Daston and Galison describe the paradigm as a pursuit to develop modes of inscription that create pure and objective visualizations without human intervention, even if just for the sake of removing dirt and imperfections~\cite{Daston2007Objectivity}. Culminating in the work of E. J. Marey, mechanical objectivity still resonates in contemporary efforts to develop canonical visualization principles based on scientific criteria. 

Mechanical objectivity in its purest ambition of tracing “nature's pencil,” however, was bound to fail due to the indispensability of narrative explanation and the ambiguous nature of the trace. Historian Carlo Ginzburg describes the interpretation of traces, clues, and symptoms as a method of conjecture rather than computation~\cite{Ginzburg1979Clues}. Philosopher Sybille Krämer locates traces “at the seam of where the meaningless becomes meaningful,” embodying meaning through material configuration rather than verbal attribution. In her understanding, traces are not found, but constructed in the act of reading: a trace is whatever is recognized as a trace~\cite{Kraemer2007Spur}. Contemporary thinkers under the umbrella of \emph{new materialism}, however, do not insist on the centrality of the human observer~\cite{VanderTuin2012New}. Distinct from both \emph{realist} (focusing on the external world) and \emph{anti-realist} (focusing on the relationships among signs) perspectives, Karen Barad's concept of \emph{agential realism} considers the human subject as a part, but not the center of an external phenomenon~\cite{Barad2007Meeting}. Avoiding any dualism between objects in the world and their representations, Barad understands a phenomenon as an ongoing process of what she describes as intra-actions rather than a fixed set of objects and their relationships.

Translated to the subject at hand, this implies that autographic visualizations are not stable artifacts whose correct interpretation is just a matter of visual literacy, but phenomena that emerge from a recipients' extensive engagement with the world and with the knowledge of others, like a hunter who learns to spot latent animal tracks that are not just invisible but non-existent for an unskilled person. Philosopher Michael Polanyi aptly describes how a complex trace can depend on theoretical concepts and language~\cite{Polanyi1998Personal}:
\begin{quote}
	Think of a medical student attending a course in the X-ray diagnosis of pulmonary diseases. He watches, in a darkened room, shadowy traces on a fluorescent screen placed against a patient's chest, and hears the radiologist commenting to his assistants, in technical language, on the significant features of these shadows. At first, the student is completely puzzled. [\dots] The experts seem to be romancing about figments of their imagination- he can see nothing that they are talking about. Then, as he goes on listening for a few weeks, looking carefully at ever-new pictures of different cases, a tentative understanding will dawn on him; he will gradually forget about the ribs and begin to see the lungs. And eventually, if he perseveres intelligently, a rich panorama of significant details will be revealed to him (p. 106) 
\end{quote}

\section{Autographic Visualization Neighbors}
Autographic visualization shares a space with other visualization models concerned with physical information displays, embedded in physical environments and contexts of action~\cite{willett_embedded_2017}. They can be seen in the tradition of ubiquitous computing and its explorations of tangible media, ambient and situated displays~\cite{Weiser1991computer,Weiser1996Designing,IshiiTangible,Wisneski1998Ambient}. 

Within the information visualization discourse, the field of data physicalization is closest to the concept of autographic visualization. Data physicalization investigates three-dimensional physical embodiments of information and their possible advantages for data communication and exploration~\cite{jansen_opportunities_2015}. Unlike autographic visualization, however, physicalization (or physical visualization) is a data-first approach. As Jansen et al.\ explain: “Traditional visualizations map data to pixels or ink, whereas physical visualizations map data to physical form”~\cite{Jansen2013Evaluating}. Data physicalization aims to take advantage of the cognitive processes involved in examining, manipulating, and constructing three-dimensional objects that may not be accessible through visual observation of two-dimensional representations. The goal of data physicalization is therefore epistemological---supporting data analysis---while autographic visualization emphasizes ontological questions such as what constitutes a datum and how it relates to the world.

Based on the Peircean concept of the index, indexical visualization presents a design space spanned by the dimensions of symbolic and causal distance~\cite{Offenhuber2015Indexical}; the former describes the amount of symbolic mediation used to transform a phenomenon into a display, the latter the number of transformations in the causal chain. Despite its short causal distance, a simple seismometer involves a high degree of symbolic mediation; its line chart can no longer be connected to the phenomenon without knowledge of the process that created it. Conversely, an ambient display that mimics the outdoor sky based on weather data would have a short symbolic distance, but a long causal distance because of the complexity of the mediating apparatus. In place of these two dimensions, the concept of qualitative displays elegantly presents a one-dimensional measure of “directness,” describing the degree of intervention by a designer~\cite{Lockton2017Exploring}. This dimension spans five different levels ranging from visual phenomena that are their own visualization to highly artificial data physicalizations at both extremes of the scale. The authors argue that visualization, so far, has been biased towards quantitative information while neglecting qualitative aspects.

Indexical visualization and qualitative displays both are motivated by a gap in existing frameworks: the neglect of the index compared to icons and symbols in the former, the neglect of qualitative information in the latter. Both emphasize the embeddedness of visualizations in the physical world~\cite{willett_embedded_2017}. Neither, however, fully capture the nature of analog visualizations of material information: Indexicality requires adhering to a semiotic framework that insists on explicating visual codes. The term qualitative display, on the other hand, seems overly broad as a descriptor of material displays. The term \emph{autographic} addresses the main difference to information design, InfoVis, and data physicalization: the self-inscribing nature of material displays, in which the designer creates the apparatus that lets traces emerge rather than explicitly defining symbolic mappings. Areas of intersection exist: for example, data visualization software that generates and displays its own data from user interaction and therefore assumes autographic qualities, or projects such as \emph{Dear Data}, when the signature of the author is considered as a trace~\cite{lupi_dear_2016}.

Autographic visualization continues the explorations into self-illustrating phenomena, first presented by Pat Hanrahan in his 2004 IEEE Vis capstone talk~\cite{Hanrahan2004Self-illustrating}. Referencing a concept from the history of scientific representation~\cite{Robin1992scientific}, Hanrahan focused on scientific experiments rather than the broader cultural field of visual practices. Autographic displays, however, are not limited to science but can be found throughout history and culture. The term autographic not only reflects the process of visualization and the role of the designer in this process but is also historically accurate, since the term was widely used during the late 19\textsuperscript{th} and early 20\textsuperscript{th}~century to describe self-inscribing mechanisms~\cite{Siegel2014Forensic}. As reflected by a Google n-gram search, the terms “autographic” and “self-registering” saw their peak in the early 20\textsuperscript{th}~century, where they show up in many patent applications for mechanical visualization devices and photographic techniques, before losing popularity later in the 20\textsuperscript{th}~century.\footnote{See~\url{https://books.google.com/ngrams/graph?content=autographic\%2Cself-registering&year_start=1800}}

\section{Autographic Design Operations} 

\begin{table}
	\caption{Overview of autographic design operations}\label{designop}\scriptsize\centering\medskip
	\begin{tabular}
		{L{2cm} L{2cm} L{3.2cm}} Objective&Operations&Description\\
		\midrule Establishing perceptual space~\cite{Latour1990Visualisation,Lynch1985Discipline} & Framing& Establishing a perceptual context to isolate a phenomenon~\cite{Bateson1972Steps,Berger2008Ways}\medskip\\
		&Constraining&Isolating a single quality by constraining other qualities~\cite{Rheinberger1997Toward,Peirce1998Essential}\medskip\\
		Tuning scale and intensity&Aggregating&Making visible by aggregating material~\cite{Rheinberger1997Toward,Latour1999Pandora's}\medskip\\
		&Separating&Making visible by separating material~\cite{Rheinberger1997Toward,Daston2007Objectivity}\medskip\\
		Trace-making& Coupling & Making visible by allowing the phenomenon to interact with another substance~\cite{Rheinberger1997Toward,Daston2007Objectivity}\medskip\\
		&Registering&Creating a persistent trace~\cite{Latour1990Visualisation,Daston2007Objectivity,Siegel2014Forensic,Goodman1968Language}\medskip\\
		Measuring and Interpretation& Annotating & Adding graphical elements to guide the interpretation~\cite{Bredekamp2015technical,Wulfingen2017Traces:}\medskip\\
		& Encoding & Adding a scale for discretizing a phenomenon~\cite{Goodman1968Language,Cole2002Suspect}\\
	\end{tabular}
\end{table}

The production of interpretable traces is facilitated by cultural techniques that involve various degrees of intervention. In the most simple case, an environmental trace presents itself to a skilled observer. At the other end of the spectrum, traces are the product of a complex experimental apparatus involving many transformations. Along this continuum, the engagement with traces can be articulated as a design process that comprises a set of operations to turn a phenomenon into encodable data. The designer has to decide which aspect of a phenomenon can be used as an indicator and proceed to apply different operations that make this indicator legible. 

The visual vocabulary of information visualization is formalized in schemata ranging from the foundational concept of visual variables to the grammar of graphics, organized by data structure and user needs~\cite{Bertin1983Semiology,Shneiderman1996eyes,Card1997structure,Wilkinson2005grammar,Wickham2010Layered}. A taxonomic approach that categorizes trace-phenomena into visual variables seems impractical and would introduce another level of symbolic representation. Instead, our approach focuses on the design operations involved in autographic design (Fig.~\ref{collage}). Table~\ref{designop} provides an overview of these operations, grouped by the kinds of transformations they achieve. Literature categorizing traces exists in domain-specific areas, from the forensic analysis of crash skid marks~\cite{Struble2013Automotive} to the identification of animal tracks~\cite{Liebenberg1990art}. But to our knowledge, there are no overarching accounts that generalize the visual operations of trace-making across disciplines. The following taxonomy is an attempt to this effect. 

The construction of the proposed autographic design space involved multiple steps. The fundamental concepts were drawn from theoretical literature, including history of science~\cite{Daston2007Objectivity}, theory of scientific representation~\cite{Latour1990Visualisation,Coopmans2014Representation}, and other perspectives on the ontology and epistemology of the trace~\cite{derrida_grammatology_2016,Kraemer2007Spur,Rheinberger2011Infra-experimentality}. We also included  professional literature from fields concerned with preparing traces, especially in medicine and the forensic sciences~\cite{Struble2013Automotive,Weizman2017Forensic,Cole2002Suspect}. 

The reviewed literature made clear that the goals of trace-making diverge across fields. While the natural sciences are interested in the generalization of the phenomenon behind the trace, forensic science is striving for individualization: finding what differentiates a particular object from all other things in the world, e.g., the gun that fired a bullet. However, despite these different objectives, the practices of identifying, preparing, and transcribing a trace share more similarities than differences. The review, therefore, focused on practices more than the underlying intent. 

The next step involved the collection and analysis of 800 examples of traces and techniques of trace-making in the broadest sense.\footnote{For reference~\url{https://www.pinterest.com/dietmaro/autographic-visualization}} These examples were examined considering the theoretical concepts identified earlier and used to reflect on these concepts. To include autographic devices such as the sundial or the Cyanometer (Fig.~\ref{collage}a), we expanded the definition of the trace from the narrow meaning of a persistent imprint~\cite{Kraemer2007Spur} to a broader definition that includes ephemeral phenomena such as shadows or sound.

The resulting design space is grouped into four sections that loosely correspond to the steps involved in trace-making: 1.\ frame the context in which the phenomenon can emerge, 2.\ adjust the intensity of a phenomenon to make it intelligible, 3.\ register the trace phenomenon and make it persistent, and 4.\ annotate and encode it into data. The steps are not strictly sequential (steps 2 and 3 are sometimes skipped), so the term “pipeline” does not seem appropriate, but they are generally traversed in one direction. The four steps synthesize and simplify operations discussed in the literature under technical image production and \emph{chains of representation}~\cite{Bredekamp2015technical,Latour1999Pandora's}.

\subsection{Establishing Perceptual Space}

In an environment saturated with latent information, the first step involves defining the space and context for the autographic visualization and thus offering a scaffolding for its reception. This problem rarely emerges in InfoVis, where charts are usually recognized as such and appear in familiar contexts such as newspapers, websites, or exhibitions. But to facilitate decoding, also traditional visualizations have to define a spatial reference system and clarify the domain covered by the data~\cite{Lynch1985Discipline}. Framing and constraining are two families of operations that focus the attention, isolate the phenomenon from its background, and offer a reference for comparison.

\paragraph{Framing}As the most fundamental autographic operation, framing circumscribes the perceptual space of the visualization. Framing guides the attention to a particular quality of a phenomenon while masking the many other qualities that are not considered relevant. Framing manipulates the context of a phenomenon without touching it. Nevertheless, framing determines how the phenomenon presents itself, shaping the qualities of the display. Framing can be illustrated through the Cyanometer, a historical device for measuring the blueness of the sky consisting of a numbered color scale with a hole at the center~\cite{Saussure1791Description}. The frame separates the color as the quality of interest from its surrounding context and simultaneously allows constructing a different context that allows comparison or measurement~(Fig.~\ref{collage}a). Framing is also a rhetorical strategy and, as such, omnipresent in information design and visualization. In communication theory, framing describes a form of meta-communication that places a message into an existing interpretative context~\cite{Bateson1972Steps}. 

\paragraph{Constraining}As a stronger form of framing, constraining involves physically manipulating the phenomenon. Constraining isolates a particular quality of a phenomenon from all others, but unlike framing it makes it observable by physically limiting the degrees of freedom of other qualities and behaviors. As an example, a glass thermometer allows a liquid to expand with temperature, but only in a single direction and by amplifying the expansion through the diameter of the tube~(Fig.~\ref{collage}b). As a sentinel species, the proverbial canary in the coal mine is constrained in a cage, so that its demise can alert miners of dangerous gases in the surrounding atmosphere. Constraining can be compared to similar interaction techniques in data visualization that control for changes in a specific variable by keeping the others constant.

\subsection{Tuning Scale and Intensity} 
After a perceptual space is established, the phenomenon might still be invisible because it is too faint or too intense, too large or too small, too fast or too slow to be perceived. The second group of autographic operations is therefore aimed at tuning the scale, speed, and intensity of a phenomenon to the gamut of human perception. Hans-Jörg Rheinberger goes as far as describing compression and dilatation as the two fundamental procedures of scientific experimentation~\cite{Rheinberger2011Infra-experimentality}. In the following, aggregation refers to operations that compress a phenomenon in space, time, and magnitude, while separation achieves the opposite. In data visualization, the two operations are equivalent, usually accomplished by tweaking visual variables. In autographic space, however, aggregation and separation are quite different in terms of operations and levels of complexity. We therefore discuss them separately.

\paragraph{Aggregation}A straightforward way to amplify the visual intensity of a material substance is to aggregate the substance over time until visual differences become apparent. Air-borne particulate matter is invisible but reveals itself in the filter of a dust mask worn over an extended period in polluted air. An example of spatial compression involves the collection of soil samples from a larger territory, which can then be organized into a grid that serves as a compact visualization of the territory~(Fig.~\ref{collage}i). Aggregating material under controlled conditions is at the core of many methods of sensing and measurement, such as the gravimetric measurement of particulate matter~(Fig.~\ref{collage}c). InfoVis methods designed to reveal patterns based on aggregation include scatter plots and heat map displays.

\paragraph{Separation}Often the opposite is necessary, untangling a material mixture and spatially separating it into its components based on their physical properties. A prism separates white light into its different wavelengths. In paper chromatography, the components of an ink blot on a piece of paper dipped into water are separated by the force of capillary action. DNA Electro-chromatography works by the same principle, separating fragments of DNA embedded in a gel driven by the force of an electric field~(Fig.~\ref{collage}d). The analytical separation of multiple correlated variables is a central task in InfoVis, addressed in methods such as scatterplot matrices, parallel coordinate displays, or faceting.

\subsection{Trace-making techniques}
Many phenomena of interest are non-visual but can be visualized through their interaction with certain substances and processes. Other visual phenomena are ephemeral; operations of marking and tracing can help to preserve traces and create a persistent record. The operations under this rubric include most analog visualization techniques, which are historically and visually linked to the contemporary languages of data visualization. 

\paragraph{Coupling} links an invisible phenomenon to a second phenomenon that serves as a visible indicator or proxy. Wind itself may be invisible, but reveals itself in the movement of grass; smoke injected into a wind tunnel serves the same purpose~(Fig.~\ref{collage}e). Many archaic skills of farming, hunting, or navigating rely on observing such proxy indicators linked with the phenomenon of interest. Coupling can involve adding a tracer substance, such as color dyes to visualize the movement of liquids. 18\textsuperscript{th}~century experimentalist Ernst Chladni visualized the shape of sound-waves by adding sand on a metal plate struck by a violin bow~(Fig.~\ref{collage}g). Mechanical instruments can be designed to respond to a phenomenon with visible changes, as illustrated by pressure gauges or meteorological instruments. By translating the phenomenon into changes in a defined visual layout, such instruments provide a bridge into the space of symbolic representation. 

\paragraph{Registering or marking} is a stronger form of coupling that aims to create a permanent trace---a cast from a footprint, the groove of a vinyl record, or the photochemical reaction in an exposed photograph. Tracer substances can also be used to create a permanent trace, including dyes to reveal structure in biological specimen, radioactive tracers used in radiology, or powder to reveal latent fingerprints. Campbell-Stokes' sunshine recorder serves its literal purpose through a spherical lens that burns a linear trace into a paper strip~(Fig.~\ref{collage}f). Registration creates not only a spatial but also a temporal record of a phenomenon. James Watt’s indicator mechanism, producing a two-dimensional line chart of steam pressure over piston displacement, became the first of many self-registering devices, including the black boxes used in aviation~\cite{Siegel2014Forensic}. Real-time data visualizations from sensor input are, to some extent, autographic visualizations.

\subsection{Measuring and Interpretation}\label{encode}
The last step in the design of autographic visualizations involves the interpretation of the trace and to make it comparable to other traces. This step can involve many different forms, from manual annotations of traces during the analysis, visual guides for non-expert viewers, to scales for encoding the trace into symbolic data.

\paragraph{Annotating} Graphical annotations are frequently found where traces and records of measurements are interpreted, creating a hybrid of graphic and autographic displays~(Fig.~\ref{collage}h). Such scribbles can themselves be seen as traces of a thought process or collective discourse. Annotations and legends can also guide the attention and support the discovery of traces where implicit framing is not sufficient. Museum displays of archeological artifacts often include abstracted representations that point out significant features of the object. In all of these examples, annotations serve largely identical purposes in information design and autographic visualization.

\paragraph{Encoding} represents the last step in the process of translating a phenomenon into data. Encoding begins by marking different conditions over time, registering observations~\cite{lupi_dear_2016}. A systematic application of such marks becomes a scale that allows encoding the phenomenon into discrete elements~(Fig.~\ref{collage}j). Fingerprints became a viable means of identification only after an efficient system of encoding their intricate appearance into a sparse sequence of symbols was developed~\cite{Cole2002Suspect}. In the operation of encoding, an analog system becomes a digital system. Nelson Goodman describes a pressure gauge as an analog system if the marks on the gauge face are used for mere orientation, but it becomes a digital system, once only the discrete intervals are considered~\cite{Goodman1968Language}

\section{Autographic Systems} 
The design operations described in the previous section form the basic vocabulary of autographic visualization. Revealing a phenomenon, however, typically requires the application of several operations, combined in an \emph{autographic system}. Most examples given earlier, when applied in practice, form autographic systems of various complexity, comprising a system of operations for framing, tuning, and recording a trace and making it measurable. A set of design operations can either be deployed in parallel to create a controlled environment to observe a phenomenon in isolation, or sequentially as a series of material transformations to make an invisible phenomenon accessible to the senses. While desire paths may appear accidental and free of design intent, complex autographic systems can be highly artificial displays that represent through analogies---as in the case of MONIAC, a hydraulic model meant to simulate economic flows~\cite{bissell_historical_2007}. In the following, three common types of autographic systems will be discussed. The list in Table~\ref{autosys}, neither exclusive nor exhaustive, includes experimental apparatuses, analog visualizations and simulations, hybrid systems using digital and analog components, and new materials with intentionally designed properties and behaviors.

\begin{table}
	\caption{Autographic systems}\label{autosys}\scriptsize\centering 
	\begin{tabular}
		{L{3cm} L{5cm}} Autographic systems&\\
		\midrule Autographic environments&Systems that combine operations to generate traces under controlled conditions\medskip\\
		Digital/physical systems&Coupling analog and digital systems\medskip\\
		Autographic materials&Encoding behavior into smart materials or synthetic organisms\\
	\end{tabular}
\end{table}

\subsection{Autographic Environments} 
\begin{figure}
	[htb] \centering 
	\includegraphics[width=3.45in]{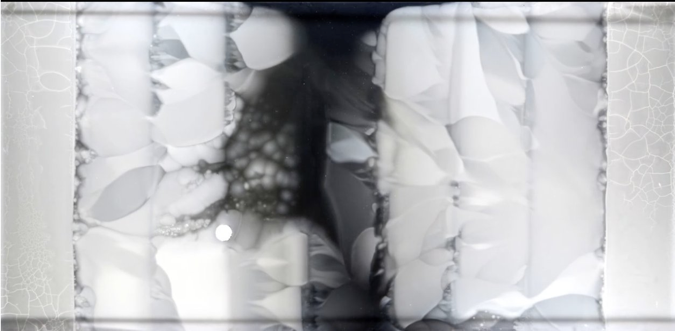} 
	\caption{Autographic environment: gel with graduated antibiotic presence to observe microbial evolution towards antibiotic tolerance~\cite{Baym2016Spatiotemporal}}\label{antibiotic}
\end{figure}

Autographic environments are environments in which a phenomenon can unfold, largely isolated from external influences. Operations such as \emph{framing, constraining, coupling,} and \emph{registering} are mobilized to turn a phenomenon into a usable trace under controlled conditions. The wind tunnel is a simulated environment isolated from its surroundings. It includes mechanisms for producing traces (e.g., through smoke or silk strings), for observing and recording them. Another example is the large bacterial growth area in Fig.~\ref{antibiotic} to study the adaptation of bacteria to antibiotic environments. Autographic environments are analog computers and visualization systems, allowing us to perform the same visualization tasks on different inputs and observe the results.

\subsection{Digital/Analog Systems} 
\begin{figure}
	[htb] \centering 
	\includegraphics[width=3.45in]{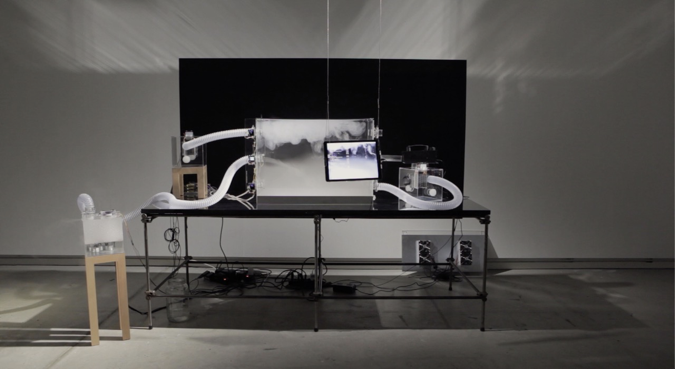} 
	\caption{Digital/physical system: Clemens Winkler. Per-forming clouds. 2018 – art project attempting to create rectangular clouds~\cite{Winkler2018Per-Forming}}\label{cloud}
\end{figure}

Digital/analog systems are a special case of autographic environments, which utilize both digital and analog forms of computation for controlling a phenomenon. The digital/analog coupling can happen in three different ways. First, the physical conditions in the autographic environment can be computationally controlled to achieve different results. Another possibility is to digitize the outputs of the apparatus and subject them to further computational analysis. The third possibility is to couple digital and analog processes in a dynamic feedback loop, creating a hybrid, autopoietic system. Since, as discussed earlier, digital sensors and circuits process material information, the line between digital and analog components, discrete logic and continuous feedback, is somewhat blurry. Fig.~\ref{cloud} shows an example of an autopoietic digital/analog system, a cloud chamber controlled by a digital algorithm aiming to sculpt the shape of generated clouds.

\subsection{Autographic Materials} 
\begin{figure}
	[htb] \centering 
	\includegraphics[width=3.45in]{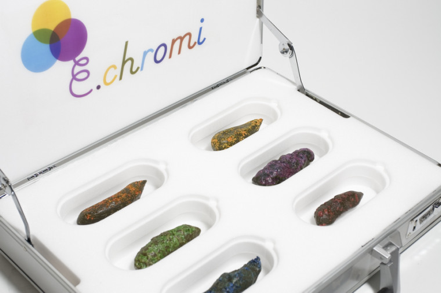} 
	\caption{Autographic material: Daisy Ginsberg, James King, E.Chromi, 2009 – color changing bacteria to detect gut diseases, a speculative design project~\cite{Ginsberg2009E-Chromi}}\label{echromi}
\end{figure}

The last group of autographic systems avoids the complexity of digital/analog apparatuses and aims to develop materials and biological organisms with truly autographic properties~\cite{telhan_designature_2016}. The concept of \emph{4D printing} investigates geometries and materials that dynamically respond to environmental changes or possess the capacity for self-assembly~\cite{tibbits_4d_2014}. Autographic materials also include work in synthetic biology with modified bacteria that react to environmental changes with visible changes of, for example, color or smell (Fig.~\ref{echromi}).

\section{Rhetoric Techniques of Autographic Visualization} 
Considering the pervasive availability of digital information and the extent to which our experience is shaped by it, are there compelling reasons, beyond nostalgia for the analog, to engage with the slow, ambiguous, and bespoke domain of material displays? The beginning of this paper presented the argument that autographic visualizations allow to experience the phenomenon behind the data and render legible the circumstances of data collection. Traces are not representations; they present themselves. This argument, however, deserves more scrutiny. Traces are often equated with incontestable evidence. As unintentional side-products of past processes and events, material traces are considered trustworthy, and their display can achieve a persuasive effect that is difficult to attain with digital representations. However, as discussed earlier, traces neither speak for themselves nor are recognized by everyone. The persuasiveness of traces is shaped by social and cultural processes, as illustrated by the slow acceptance of fingerprinting and DNA identification~\cite{Cole2002Suspect}. Traces require interpretation, and their interpretation relies on assumptions and often speculation. The most apparent patterns are often misleading. Interpretation, therefore, requires a rhetoric scaffolding for integrating material displays into a framing argument. The following section discusses examples of autographic design principles as they are used in practice. In all of these cases, material traces are used as rhetorical devices; as visual arguments that support specific claims. 

\subsection{Evidentiary Aesthetics in Citizen Science}
Tufte’s imperative of information visualization to “above all else show the data” becomes “above all else move closer to the phenomenon” in autographic visualization. The tactic to enable the sensory experience of causality is often found in the domain of citizen science~\cite{Snyder2017Vernacular}. Especially groups who investigate issues of environmental pollution are often met with skepticism of the data they collect—whether their methods are rigorous, their instruments accurate, or their biases reflected in data collection. Many grassroots scientists lack institutional affiliations and scientific credentials, making their data sets often seem less trustworthy in public perception. In such an environment, the evidentiary aesthetics of a raw trace can be a more effective way to generate trust than a well-designed chart using canonical visualization principles.

\begin{figure}
	[ht] \centering 
	\includegraphics[width=3.45in]{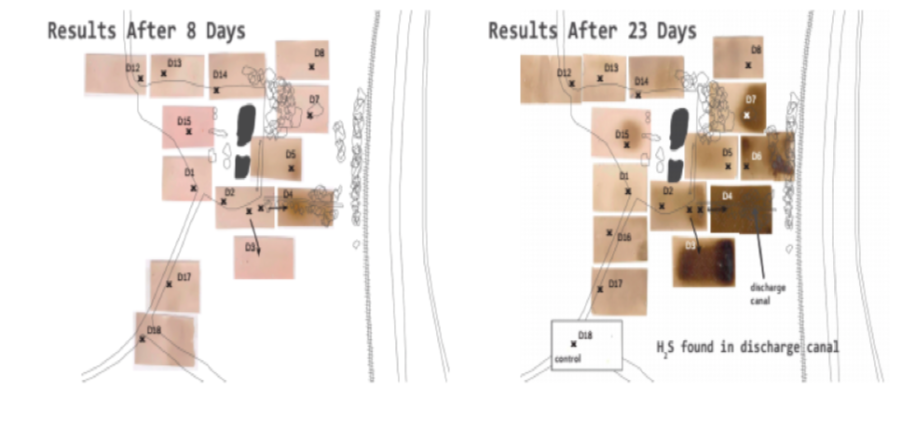} 
	\caption{Map of photo strips tarnished by H2S emanation on the field site, Public Lab. Registering, annotating~\cite{Wylie2017Materializing}}\label{h2s}
\end{figure}

The Public Lab is a grassroots science collective investigating environmental pollution and the social impacts of oil spills and fracking; issues that in their view do not receive enough attention by environmental agencies~\cite{Wylie2017Materializing}. The group has developed a DIY method using photo paper to detect harmful hydrogen sulfide gas emanating from the ground in proximity to fracking sites. The maps presenting their results incorporate an arrangement of small pieces of the original, stained photo papers over an abstracted representation of the landscape (Fig.~\ref{h2s}). The grid of samples not only shows a visual pattern similar to a heat map indicating the locations of highest exposure, it also suggests physical circumstances to the uninitiated viewer: that a chemical reaction that stains photo paper is associated with these places, implying the papers were actually exposed at the indicated locations (a rhetorical claim that cannot be verified using the map). In combination with an encoded data set, the trace map serves as an illustration and justification of the method. But the autographic map is not only directed outwards towards a skeptical audience but also inwards at the own collaborators. The modes of data collection are often participatory and depend on the engagement of volunteers and members of the affected community. To this end, data collection is staged as a public experiment; the physical traces serve as rhetorical devices to make the nature of pollution tangible for collaborators, and the results of their voluntary efforts visible. It may be a lucky accident that the pollutant causes an noticeable and reproducible trace to visualize environmental harms. Autographic design involves discovering such opportunities and building an explanatory framework around a suitable indicator. The first rhetoric strategy involves choosing a presentation that suggests immediacy and direct causal connection over more abstract representations. In the case of Public Lab’s grassroots science, the presentation of causality is not just concerned with data veracity, but, more importantly, with the methods and practices of its researchers. 

\subsection{Performative Mapping and Annotated Walkthroughs}\label{performative}
The second rhetoric strategy involves guiding the audience through the causal chain, helping them to “connect the dots” and perform the analysis themselves. Traces imply causality, but the immediate cause is absent from the trace—what is left is an imprint. Annotated walkthroughs put the traces that are considered relevant next each other and allow the recipient to explore the latent connections. 

The example to illustrate this approach is based on digital information---satellite images, video footage, and other media formats read through a material-forensic lens, highlighting the indexical and material “residues” in digital data. In recent years, an active community of conflict mappers and amateur forensic experts has emerged who analyze and compare social media content from conflict regions and gained prominence during the Syrian civil war~\cite{Kurgan2017Conflict,Weizman2017Forensic}. In the first major military conflict in which social media played a decisive role, all adversaries made extensive use of platforms such as YouTube or Twitter to disseminate footage from the frontlines recorded by drones, smartphones, and body-cams. Social media not only served propagandistic purposes but also as a backchannel for reporting military success to foreign donors, which in some cases even involved staging fake battles~\cite{Atwan2015Islamic}.

\begin{figure}
	[htb] \centering 
	\includegraphics[width=3in]{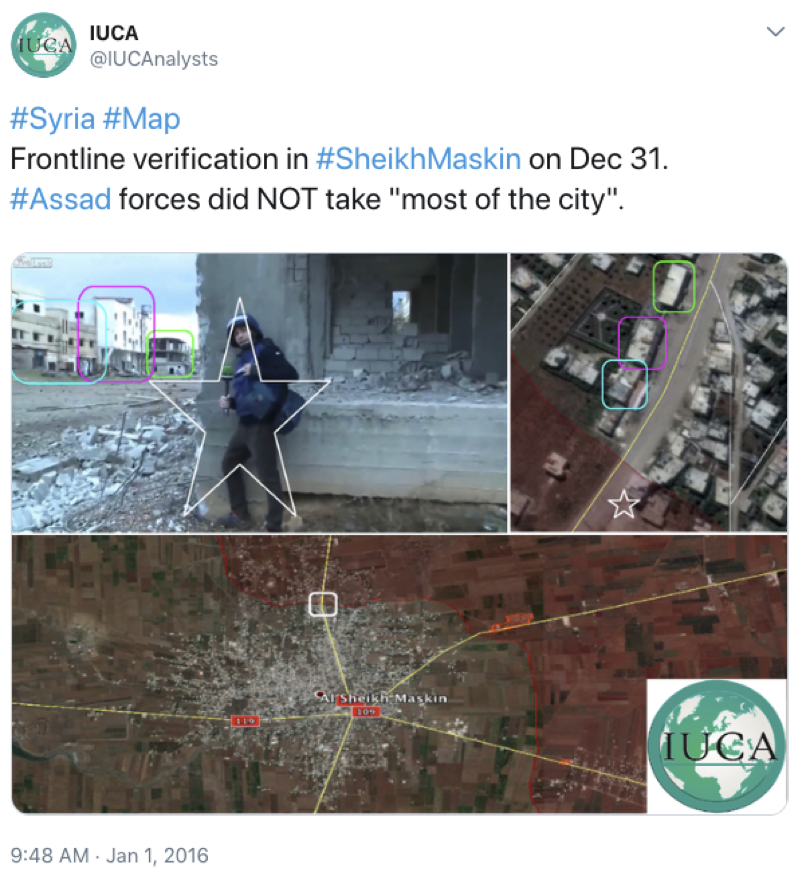} 
	\caption{Amateur visual forensics, Institute for United Conflict Analysts – framing, annotating (IUCA)~\cite{IUCA2016Syria}}\label{syria}
\end{figure}

The community of conflict mappers took it upon themselves to verify claims of the warring parties by geo-referencing buildings and locations shown in the videos and placing the events into a temporal sequence in order track the shifting frontlines~\cite{Offenhuber2017Maps}. As in the case of the grassroots scientists, the conflict mappers are not represented by certified institutions, and therefore choose their visual displays to address questions of credibility. For their frontline analysis, conflict mappers have created a specific format of display. It does not synthesize information from individual sources into a consistent cartographic language but arranges snippets from the raw sources. Elements in this tableau are annotated with simple shapes, indicating the same building or location from various angles in smartphone footage and satellite images. In the example featured in Fig.~\ref{syria}, conflict mappers countered the claim of the Syrian military to have captured a particular city by demonstrating that the shown location is not inside the city, but rather on its outskirts. To interpret these displays, the recipient is forced to do the work of the cartographer, judge the likelihood of whether the highlighted elements are the same building or whether they are shot at the same time of day. This rhetoric strategy has been described as non-representational or performative cartography~\cite{Kitchin2007Rethinking}. The same strategy of guiding the viewer by juxtaposing elements, implying connection through spatial proximity, and highlighting relevant aspects can be equally applied to material traces, is used in museum displays of archaeological artifacts. The displays of the conflict mappers try to persuade not by encoding information into visual variables, but by emphasizing the authenticity of the sources and inviting the viewers to “see for themselves.” Their use of publicly available tools such as Google Earth without bothering to modify their visual defaults underscores this invitation as if to say that anyone can conduct this investigation and arrive at the same conclusion. 

\subsection{Sensory Accountability---Exploring Data Materiality}
The third rhetoric strategy contextualizes digital data and computational models with material displays to explore their grounding in physical reality. A considerable number of artists have taken on visualizing the materiality of climate change and environmental pollution through situated and material displays. Proxy data sources play, implicitly or explicitly, a major role in this genre: from visceral explorations of the rich material qualities of ice core samples and arctic ice to the instrumentalization of bioindicators and sentinel species---plants and animals that are especially sensitive to particular conditions and can serve as environmental sensors. Many of these projects not only comment on environmental phenomena such as climate change, but also on the aesthetic and ontological dimensions of scientific measurement: what is it that is measured, which qualities are captured or overlooked, and what are the political underpinnings of how a problem is articulated and operationalized. 

\begin{figure}
	[htb] \centering 
	\includegraphics[width=3.45in]{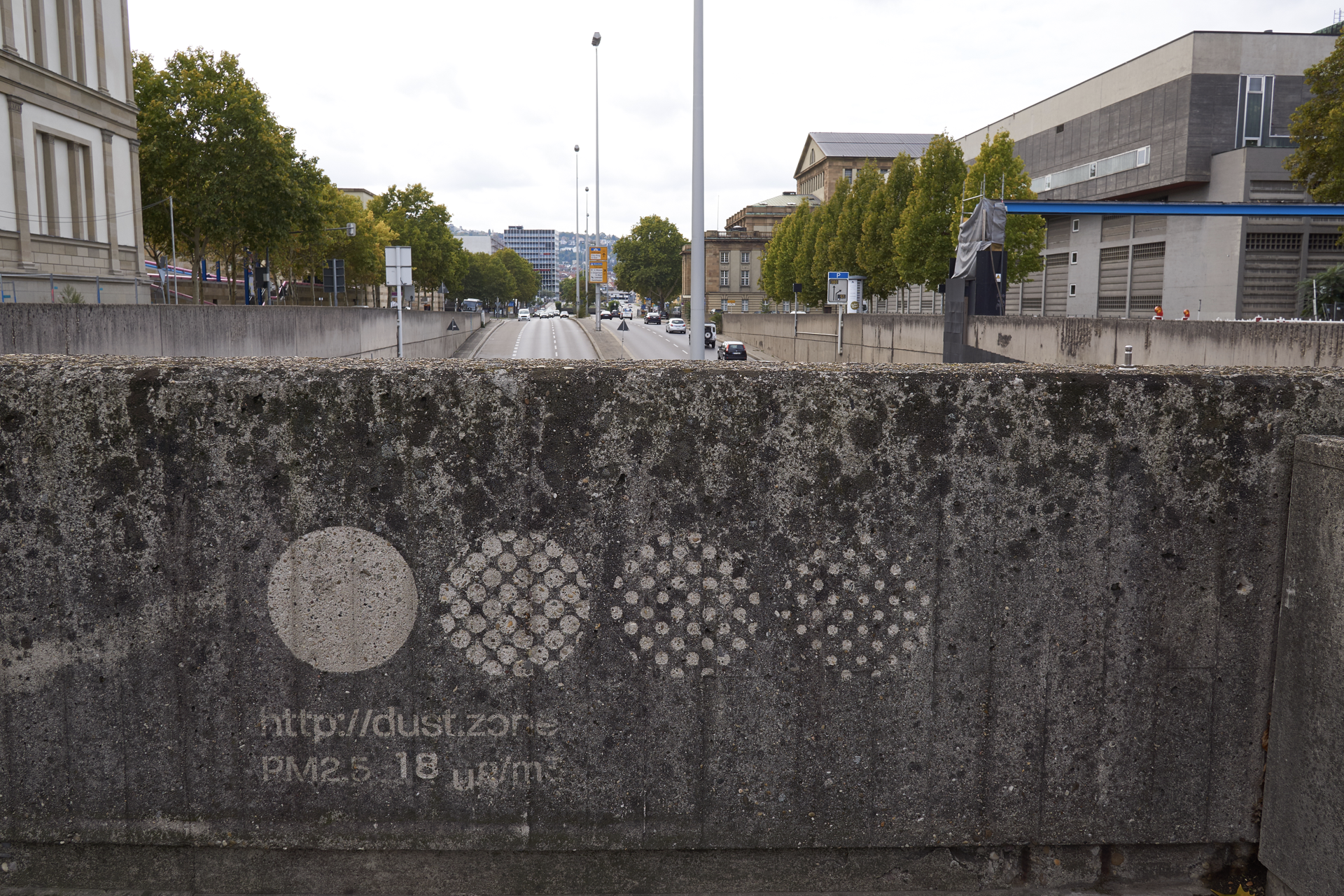} 
	\caption{D. Offenhuber, Staubmarke – reverse graffiti washed into concrete, calling attention to air pollution, aggregating, encoding~\cite{Offenhuber2018Dust}}\label{staub}
\end{figure}

In public controversies around air pollution stemming from particulate matter and ozone, the discourse among actors with conflicting positions often quickly converges on technicalities of measurement such as appropriate threshold values and exposure times. The work of grassroots science initiatives thus has a strong political dimension: challenging conventional protocols of measurement by revealing what is not shown by them. Their perspective of data collection values richness and completeness over accuracy; aims to capture the implications of pollution on the lives of individuals and communities even if the accuracy of the cheap sensors used in these projects is “good enough” rather than perfect. In this context, material displays can be used to call attention to the basic assumptions of environmental sensing and their political implications. We use the term “sensory accountability” to describe approaches that call attention to the complex material qualities of a phenomenon that is often reduced to a single quantitative dimension. As an example, the project “Staubmarke” (Fig.~\ref{staub}) visualizes air pollution by applying visual markers on urban surfaces to make accumulations of particulate matter legible. These markers are executed as \emph{reverse graffiti}, which are based on selectively cleaning dirty surfaces rather than applying paint. With ongoing pollution, the markers will fade over time, starting with the most delicate textures in the pattern. The markers were applied at locations also surveyed by the sensors installed by a citizen sensing initiative, allowing the comparison between data values and physical appearance of the markers. Sensory accountability is a critical inquiry into sensing methods: what is captured by the sensor, what is ignored, and how do the recorded values correspond with sensory phenomena. Highlighting changes in the physical environment and contextualizing these changes with corresponding data offers an interesting space for visualization.

\section{Discussion---Reviving the Public Experiment}
Physical expressions of data currently enjoy burgeoning interest, from the popularity of data physicalization in design and education to the wide range of artistic projects focusing on data and the environment. One can make the argument that this is not merely a short-lived design trend, but an expression of a new sensibility for the relationship between digital data and the world around us. The current fascination with materiality in design and the humanities coincides with renewed critiques of the central role of data in society and a discomfort with the explicit encoding of the world into symbolic categories. Fields such as critical data studies probe the assumptions behind data collection, the politics of categorization, and the opacity of algorithmic decision systems. Research on the interpretability of machine learning and algorithmic decision-making models is currently thriving in information science and information visualization. Autographic visualization also resonates with the recent resurgence of analog computing, advances in microfluidics and synthetic biology~\cite{Ulmann2013Analog,Whitesides2006origins}. The disconnect between science and public opinion in the issue of climate change and the role of climate data as a proxy-site to negotiate political positions further demonstrate the need to probe the material foundations of data generation. Disinformation and the phenomenon fake news also provide reasons why it is crucial to investigate data generation through a forensic-material lens, based on the premise that “no two things in the physical world are ever exactly alike”~\cite{Kirschenbaum2008Mechanisms:}, which extends to digital camera sensors and hard drives. We argue that these issues also justify a shift in the agenda of visualization from the patterns inside data to the conditions of data generation. Surely, misleading patterns and false narratives can be sufficiently addressed with better explanations and more nuanced visualizations that are contextualized with scientific arguments. But the original impetus of visualization has always been to show rather than tell, to produce images that say more than a 1000 words. Autographic visualization continues this project by extending information visualization beyond the space of symbolic encodings into the spaces where data take shape. The practices of grassroots science offer an interesting model that connects to the history of public experiments during the enlightenment period, where natural phenomena were explained through spectacular public demonstrations~\cite{Nieto-Galan2016Science}.

\section{Limitations of Autographic Visualization}
Many trace-phenomena have charismatic qualities; they fascinate and invite exploration while remaining elusive. As Michelangelo Antonioni’s movie “Blow Up” illustrates, as one gets closer to a trace, the meaning seems to disappear: through successive magnification, a candid photograph reveals a murder scene but eventually dissolves into ambiguous patterns.

One has to avoid a naïve empiricism that uncritically elevates trace-reading over theoretical inquiry as a source of knowledge. Traces often seem to inspire attributions of meaning even when there is none. With regards to trace-reading, science and superstition are often uncomfortably close~\cite{Ginzburg1979Clues,butz_superstition_2007,huxley_method_1880}. As previous sections have pointed out, the charisma of traces can be exploited for rhetorical purposes by selectively curating, framing, and guiding the process of interpretation, and autographic visualization is not different from any other visual practice in this regard.

The strongest limitations are encountered on the practical level. Compared to data visualization, the creation of autographic visualizations is slow and limited by the available material. Autographic visualizations lack the agility, versatility, and potential scale of computational analysis. Where data visualization is limited by the gap between data and physical phenomenon, autographic visualization cannot compete with the full scope of computational possibilities. 

\section{Conclusion}
This paper introduces the concept of autographic visualization, which examines the discovery and preparation of material traces. It offers a preliminary systematization of the design space of autographic display including its design operations, their combination into composite autographic systems, and the rhetorical strategies of presenting material traces. Autographic visualization considers the structures in the physical world as a form of data and thus serves as a speculative counter-model to data visualization, which is limited to the space of symbolic representation. 

In the tradition of Marey's \emph{graphical method}, which encompasses both the production of traces and the graphical display of data, we argue that InfoVis and autographic visualization can complement each other. Expanding the scope of visualization beyond symbolic data would open fertile areas for research, addressing questions such as: how do we see traces, and how do these perceptions relate to individual knowledge and skills? J.J. Gibson’s theory of affordances~\cite{Gibson1979ecological} has been foundational for the entire field of HCI but has been so far mostly operationalized from a functional perspective, without further attention to the phenomenology of affordances. In this context, also the aesthetics of experience and its relationship to knowledge construction (\emph{aisthesis}) deserve attention.

Beyond academic questions, what is the place of autographic approaches in visualization practice? Artists and citizen scientists provide examples that show how the immediacy, directness, and richness of material information can be utilized. Material traces serve as visual means of evidence construction: public experiments turn data collection from a bureaucratic exercise into a sensory experience of causality; physical data proxies make abstract climate models and their predictions relatable and observable.

Autographic visualization aims not just to bridge the gap between data and phenomenon, but also the one between observer and display. Designers can no longer blindly rely on normative conventions on how to visualize data for an idealized, data-literate audience. In the space of material information, the observer becomes an experimenter, having to actively construct evidence by connecting the dots. Autographic visualization is therefore a critical practice in the sense that it de-naturalizes the concept of data and its underlying assumptions.

\acknowledgments{My gratitude goes to my long-term collaborators Orkan Telhan and Gerhard Dirmoser for their ideas and inspiring discussions. I also would like to thank the reviewers for their close attention and helpful feedback. }

\bibliographystyle{abbrv}

\bibliography{ref} 
\end{document}